\documentclass[prl,reprint,superscriptaddress,showpacs,showkeys,preprintnumbers,nofootinbib]{revtex4-1}
\usepackage[utf8]{inputenc}
\usepackage{amsmath,amsfonts,amssymb,amsbsy}
\usepackage{mathrsfs,latexsym}
\usepackage{graphicx}
\usepackage{color}
\usepackage{hyperref}
\usepackage{macros_alpha}
\usepackage{dcolumn}
\usepackage{booktabs}

\usepackage{defs}

\begin{document}
\pdfoutput=1

\preprint{DESY 16-053}
\preprint{IFT-UAM/CSIC-16-029}
\preprint{CERN-TH-2016-060}
\preprint{TCDMATH 16-04}
\preprint{WUB/16-00}
\title{%
Determination of the QCD  $\Lambda$-parameter \\ and the accuracy of
   perturbation theory at high energies}
\date{\today}
\collaboration{ALPHA}

\author{Mattia~Dalla~Brida}
\affiliation{John von Neumann Institute for Computing (NIC), DESY, Platanenallee~6, 15738 Zeuthen, Germany}
\author{Patrick~Fritzsch} %
\affiliation{Instituto de F\'{\i}sica Te{\'o}rica UAM/CSIC, Universidad Aut{\'o}noma de Madrid,\\
C/ Nicol{\'a}s Cabrera 13-15, Cantoblanco, Madrid 28049, Spain}%
\author{Tomasz~Korzec} %
\affiliation{Department of Physics, Bergische Universit\"at Wuppertal, Gau\ss str. 20,
42119 Wuppertal, Germany}%
\author{Alberto~Ramos} 
\affiliation{CERN, Theory Division, Geneva, Switzerland}
\author{Stefan~Sint}%
\affiliation{School of Mathematics, Trinity College Dublin, Dublin 2, Ireland}%
\author{Rainer~Sommer}
\affiliation{John von Neumann Institute for Computing (NIC), DESY, Platanenallee~6, 15738 Zeuthen, Germany}
\affiliation{Institut~f\"ur~Physik, Humboldt-Universit\"at~zu~Berlin, Newtonstr.~15, 12489~Berlin, Germany}

\begin{abstract}
We discuss the determination of the strong coupling 
$\alpha_\msbar(m^{}_\mathrm{Z})$ or equivalently the QCD
$\Lambda$-parameter. Its determination
requires the use of perturbation theory in 
$\alpha_s(\mu)$ in some scheme, $s$, and at some energy scale $\mu$. 
The higher the scale $\mu$ the more accurate  perturbation theory becomes, owing to asymptotic freedom.
As one step in our computation of the $\Lambda$-parameter in three-flavor QCD, 
we perform lattice computations in a scheme
which allows us to non-perturbatively reach very high energies, corresponding to $\alpha_s = 0.1$ and below. 
We find that {  (continuum)} perturbation theory is very accurate there, yielding a three percent error in the $\Lambda$-parameter,  
while data around $\alpha_s \approx 0.2$ is clearly insufficient to quote such a precision. It is
important to realize that these findings are expected to be generic,
as our scheme has advantageous properties regarding the applicability
of perturbation theory.

\end{abstract}



\maketitle

\section{Introduction}

The fundamental parameter of the strong interactions,
the coupling $\alpha_\msbar(\mu)=\gbar^2_\msbar(\mu)/(4\pi)$, is an essential input parameter for theory predictions 
of high energy processes in particular
the physics at the LHC \cite{Dittmaier:2012vm,Heinemeyer:2013tqa,Accardi:2016ndt}.
Conventionally the running
 $\alpha_\msbar(\mu)$ is quoted at
the electroweak scale, $\mu=m^{}_\mathrm{Z}$. There 
the coupling is weak, $\alpha=\rmO(1/10)$, perturbation theory
(PT) is usually accurate.
In particular $\alpha_\msbar(m^{}_\mathrm{Z})$ is essentially 
equivalent
to the renormalization group invariant $\Lambda$-parameter
\begin{eqnarray}
    \Lambda^{}_\msbar &=& \varphi^{}_\msbar(\gbar^{}_\msbar(\mu))\, \times\, \mu\,, 
\end{eqnarray}
because the function 
\begin{eqnarray}
    \varphi_s(\gbar_s) &=& ( b_0 \gbar_s^2 )^{-b_1/(2b_0^2)} 
        \rme^{-1/(2b_0 \gbar_s^2)} \label{e:phig}  \\
     && \times \exp\left\{-\int\limits_0^{\gbar_s} \rmd x\ 
        \left[\frac{1}{\beta_s(x)} 
             +\frac{1}{b_0x^3} - \frac{b_1}{b_0^2x} \right] \right\} \, 
        \nonumber                                
\end{eqnarray}
is known precisely by replacing the renormalization group $\beta$-function by its  perturbative expansion 
$\betapert_s(g)=-g^3 \sum_{n=0}^{\lb-1} b_{n,s} g^{2n}$;
in the $\msbar$-scheme $\betapert_\msbar(g)$
is known up to $\lb=4$ loops \cite{MS:4loop1,Czakon:2004bu}. 

At lower energies,  $\mu \ll m^{}_\mathrm{Z}$, the 
perturbative uncertainty in approximating 
$\beta_s\approx\betapert_s$
in \eq{e:phig} is generally not negligible. It is 
$\Delta \Lambda_s/\Lambda_s = \Delta \varphi_s/\varphi_s = 
c_\lb\alpha^{\lb-1} + \ldots$ with coefficients
$c_\lb$, which are, for
$\lb\leq 4$, of order one in the $\msbar$ scheme and expected to be
so in ``good'' schemes in general.

While the $\msbar$ scheme makes sense only perturbatively,
physical schemes defined beyond the perturbative expansion  
are easily derived from short-distance QCD
observables $\obs_s(\mu) = c_1^s \gbar^2_\msbar(\mu) + \rmO(\gbar^4_\msbar(\mu))$  via
\begin{eqnarray}
  \gbar^2_s(\mu) \equiv \obs_s(\mu) / c_1^s = 
   \gbar^2_\msbar(\mu) + \rmO(\gbar^4_\msbar(\mu))\,.
   \label{e:gengdef}
\end{eqnarray}
It is clear that high energies $\mu$ (small $\alpha_s$)
and at least $\lb=3$ are needed
if one aims for a precision determination of 
$\alpha_\msbar(m^{}_\mathrm{Z})$. Replacing high energy by just
a larger $\lb$ is dangerous because the perturbative expansion
is only asymptotic, not convergent, and non-perturbative ``corrections'' can be large. In particular, whether one has lost control is difficult to detect because our knowledge of non-perturbative physics 
is very incomplete.
Thus it is a challenge to
reach an accuracy of a few percent in $\Lambda_\msbar$ equivalent to 
sub-percent accuracy in $\alpha_\msbar(m^{}_\mathrm{Z})$.

Unfortunately, the determinations which quote the smallest
uncertainties do  typically not come from observables at large 
$\mu$ and uncertainties are dominated by systematics such
as unknown higher order perturbative and non-perturbative
terms. 
Both the Particle Data Group \cite{Agashe:2014kda} 
and the Flavour Lattice Averaging Group \cite{Aoki:2013ldr} 
are therefore not just taking weighted averages of the individual
determinations to arrive at their world averages. 

Here we consider a family of observables (schemes) 
where lattice simulations allow one {\em simultaneously to reach 
high precision and high energy before using} PT. 
Then PT at $\mu=\rmO(m^{}_\mathrm{Z})$ can be employed with 
confidence. In addition one can check its applicability 
at lower scales. {  The crucial feature enabling
the study of PT at high energy with continuum extrapolated
non-perturbative lattice results is that we use a finite volume renormalization scheme \cite{Luscher:1991wu,Luscher:1993gh}. 
QCD is considered inside a small volume of linear extent $L$
with boundary conditions and observables which do not contain
any other scale. 
Details will be presented below. 
The renormalization scale then is
\begin{eqnarray}
    \mu =1/L\,,    
\end{eqnarray}
and the continuum limit of lattice simulation results is easily
taken for $L/a \gg 1$, with modestly sized lattices.
}
This is the strategy of the ALPHA collaboration
but so far it was mostly restricted to unphysical models
with an insufficient number of quark flavors~\cite{Luscher:1993gh,Capitani:1998mq,DellaMorte:2004bc}.
For the interesting case of $\nf=3$ QCD,
the strategy was applied by the CP-PACS collaboration
\cite{Aoki:2009tf}. 
We now have very precise results for $\nf=3$ which allow us
to see important details previously hidden by uncertainties 
(see also \cite{Tekin:2010mm}).

In this letter
we discuss the most essential step: the accuracy of PT for couplings $\alpha \lesssim 0.2$ 
and our resulting precision for $\Lambda$. We will see that it is crucial to non-perturbatively reach $\alpha \approx 0.1$ 
to have confidence
in PT at the  3-4 percent level in $\Lambda$. 
{ 
On the other hand at $\alpha \geq 0.15$ and using the three-loop 
beta-function, one of our schemes
($\nu=-1/2$) shows a 10\% systematic error in $\Lambda$. 
This is not a statistical fluctuation as we will
demonstrate by \eq{e:vbar3eff}.  

Given that a priori our scheme has favorable properties
for PT and that
other tests of perturbation theory with similar precision 
and similarly small $\alpha$ are presently not available,
our result gives reason for concern
in determinations of $\alpha_\msbar(m^{}_\mathrm{Z})$ from
$\mu$ values of a few GeV. This kind of lack of accuracy of PT 
may be one of the sources of the spread of results 
reviewed in \cite{Agashe:2014kda}.
}

\section{The SF scheme}
{  Our scheme is based on the so-called
Schr\"odinger functional (SF) \cite{Luscher:1992an}. 
There are several introductory texts on
the topic with emphasis on different aspects, from
the general field theoretic concept  \cite{Luscher:1998pe}
to detailed descriptions \cite{Sommer:1997xw,Sommer:2006sj}
and a review of concepts, history and recent results \cite{Sommer:2015kza}.
Here we just summarise those aspects which
are needed to judge our findings below.
Dirichlet boundary conditions are imposed in Euclidean time,
\begin{eqnarray}
    A_k(x)|_{x_0=0} = C_k\,,
    \quad
    A_k(x)|_{x_0=L} = C_k'\,,    
\end{eqnarray}
for $k=1,2,3$.
The gauge potentials $A_\mu$ are taken periodic in space with period $L$.\footnote{
Quark fields are included as described in \cite{Sint:1993un}.
Their periodicity angle, $\theta$, introduced in 
\cite{Sint:1995ch}, is set to $\theta=\pi/5$.}
The six dimensionless matrices 
\begin{align}
LC_k &= i \,{\rm diag}\big( \eta-\tfrac{\pi}{3}, \eta(\nu-\tfrac{1}{2}), -\eta(\nu+\tfrac{1}{2}) +\tfrac{\pi}{3} \big) \,,
\nonumber \\
LC^\prime_k &= i \,{\rm diag}\big( -(\eta+\pi), \eta(\nu+\tfrac{1}{2}) +\tfrac{\pi}{3},-\eta(\nu-\tfrac{1}{2})+\tfrac{2\pi}{3} \big)\,,
\nonumber
\end{align}
just depend on the two real parameters $\eta,\nu$, which multiply the Abelian generators of SU(3).

With these boundary conditions the field which minimizes the action is
unique up to gauge equivalence~\cite{Luscher:1993gh} and denoted by $A_\mu = B_\mu^{\rm class}$. In the temporal gauge, $B_0=0$, it is given by 
$B_k^\mathrm{class}(x) = C_k + (C_k'-C_k)x_0/L $ and  corresponds to a constant color electric field. 
A family of couplings~\cite{Sint:2012ae}, $\bar{g}_\nu$, is then obtained by
taking $1/\obs_\nu$ in \eq{e:gengdef} to be the $\eta$-derivative of the effective action.}
This yields a simple path integral expectation value,
\begin{eqnarray}
  \langle \partial_\eta S|_{\eta=0} \rangle = \frac{12\pi}{\gbar^2_\nu}\,,
\end{eqnarray}
which is well suited for a Monte Carlo evaluation in the 
latticised theory. Small fluctuations around the background field
generate the non-trivial orders in PT.
It is worth pointing out that the whole one-parameter family of couplings
can be obtained from numerical simulations at $\nu=0$,
as the $\nu$-dependence is analytically known,
\begin{equation}
  \label{eq:gbarnu}
   \frac{1}{\gbar^2_\nu}  =  \frac{1}{\gbar^2} - \nu \vbar\,,
\end{equation}
in terms of the $\nu=0$ observables $\gbar^2 \equiv \gbar^2_{\nu=0}$ and $\vbar$.

Advantageous properties of these couplings are:
\begin{enumerate}
  \setlength\itemsep{-0.2em}
\item $\Delta_\mathrm{stat} \gbarnu^2 = s(a/L) \gbarnu^4 +\rmO(\gbarnu^6)$,
    for $\Delta_\mathrm{stat}$ the statistical error at a given
    length of the Monte Carlo sample.
    This property makes it possible to maintain high precision at high energy. 

\item The typical $\sim \mu^{-1},\mu^{-2}$ renormalon contributions~\cite{Beneke:1998ui} are absent since the finite volume
    provides an infrared momentum cutoff. 
    Instead, the leading known non-perturbative contribution is due to a 
    secondary stationary point of the action \cite{SFcoupinpreparation} at 
    $g_0^2\,[S(B^\mathrm{sec}) - S(B^\mathrm{class})] = 32.9$.
    It generates corrections to PT of order
\begin{eqnarray}
      \exp(-{2.62}/\alpha) 
      \sim (\Lambda/\mu)^{3.8}\,,     
\end{eqnarray} 
    which evaluates to $\rmO(10^{-6})$ for $\alpha=0.2$.
    At such values of $\alpha$, fields with non-zero 
    topology are {\em even further} suppressed given that
    $g_0^2\,[S_{|Q|\geq 1} - S(B^\mathrm{class})] \geq 6\pi^2$~\cite{Luscher:1992an,Luscher:1993gh}.
\cbla    
\item The $\beta$-function is known including its three-loop term,
    \begin{eqnarray}
      (4\pi)^3 \times b_{2,\nu}= -0.06(3) - \nu \times 1.26\,,
      \;(\nf=3)
    \end{eqnarray}    
    and for reasonable values of 
    $\nu$ 
    the three-loop term is of order one as it is in the $\msbar$
    scheme.
\item As we will see discretisation effects are very small; 
at tree-level of perturbation theory they are $\rmO((a/L)^4)$. They are known  to two-loop order in PT~\cite{Bode:1999sm} and we 
  can subtract those pieces~\cite{deDivitiis:1994yz}.
\end{enumerate}

The downside of the SF scheme is that the coefficient 
$s(a/L)$ diverges like $(L/a)^{1/2+z}$ for large $L/a$ and
is not that small in general. Here $z$ is the dynamical critical
exponent of the algorithm while the $1/2$ in the exponent is 
due to the variance of the observable \cite{deDivitiis:1994yz}.
High statistics is needed
and our computation is limited to $L/a\leq 24$.
A second issue is 
the acceleration of the approach to the continuum limit through
Symanzik improvement.
With our Dirichlet boundary conditions 
the Symanzik effective Lagrangian contains 
terms located at the time-boundaries. These
are responsible for $\Oa$ effects. We cancel them by corresponding
improvement terms with coefficients $\ct$ and $\cttilde$ known only  
in PT, see below.

\section{Step scaling functions and $\Lambda$-parameter}

The non-perturbative energy dependence of finite
volume couplings is constructed from the step scaling function \cite{Luscher:1991wu}
\begin{eqnarray}
  \sigma_\nu(u) =
  \left. \gbarnu^2(1/(2L))
  \right|_{\gbarnu^2(1/L)=u,m=0} \,,
\end{eqnarray}
where $m=0$ ensures the quark mass independence of the scheme~\cite{Weinberg:1951ss}.
The \SSF\ corresponds to a discrete version of the $\beta$-function and is computed as the continuum limit $a/L\to 0$ of its
lattice approximants $\Sigma_\nu(u,a/L)$. The conditions $\gbarnu^2(1/L)=u$ and $m=0$ then refer
to a $(L/a)^4$ lattice and fix the  bare coupling and bare quark mass of the theory. $\gbarnu^2(1/(2L))$ is to be
evaluated for the same bare parameters on a $(2L/a)^4$ lattice.

We will use the $\nu=0$ scheme for a reference, dropping the index
$\nu$. The scale $\Lswi$ is defined by a value $u_0$ and the
condition 
\begin{equation}
  \label{eq:gL0}
\gbar^2(1/\Lswi)=u_0\,.  
\end{equation}
The solution of the implicit equation 
\begin{eqnarray}
  u_k = \sigma(u_{k+1}),
\end{eqnarray}
for $u_{k+1}$, $k=0,1,\ldots$ gives a series of couplings
$u_k=\gbar^2(2^k/L_0)$. 
With a few steps one reaches $\mu =1/L_n= 2^n/L_0 = \rmO(m^{}_\mathrm{Z})$
and the perturbative $\varphi$ at this high scale 
will give a good approximation to $L_0\Lambda$
\begin{equation}
  L_0\Lambda = 2^n \varphi(\sqrt{u_n})\,. \label{e:LLpert}
\end{equation}
Note that thanks to \eq{eq:gbarnu} and the exact relation between
$\Lambda$-parameters~\cite{Luscher:1993gh,Sint:1995ch}
\begin{eqnarray}
  r_\nu = \Lambda/\Lambda_\nu = \rme^{-\nu \times 1.25516}\,,
\end{eqnarray}
the same combination $L_0\Lambda$ can be obtained in any scheme with
$\nu\ne0$. Whether different values of $\nu$, number of steps ($n$)
and perturbative orders ($\lb$) give consistent results is an 
excellent way to test the reliability of perturbation theory.

\section{Simulations}

We used the standard Wilson plaquette action and three massless $\Oa$-improved \cite{Luscher:1996sc,Yamada:2004ja}
quarks simulated by a variant of the
\texttt{openQCD}  code \cite{Luscher:2012av,openqcd:2013}. 
At eight couplings $\gbar^2(1/L)$ in the range 
$1.11 -    2.02$, we simulated pairs of lattices
$L/a,2L/a$ with $L/a=4,6,8$ and at three couplings we
also included $L/a=12$. 

Between 80k and 300k  
independent Monte Carlo measurements were taken on each lattice.
As we have already  noted, non-trivial topology is very suppressed
in these small volumes \cite{Luscher:1982wf}. Therefore topology freezing \cite{DelDebbio:2002xa,Schaefer:2010hu} is irrelevant here.

A critical issue for any lattice computation is the 
removal of discretization effects. In preparation 
of our continuum extrapolations we apply 
both Symanzik improvement of the action and
perturbative improvement of the \SSF~\cite{deDivitiis:1994yz}.
In comparison to earlier work, we here propagate the 
estimated uncertainty of those $\Oa$ improvement coefficients which
are only known perturbatively into the errors of the step scaling
functions. They can then be assumed to be free of linear $a$-effects within
their errors. Details are found in the supplementary material attached 
at the end of the paper.

\section{Continuum extrapolations and results}

\begin{figure}[t]
   \includegraphics*[width=\linewidth]{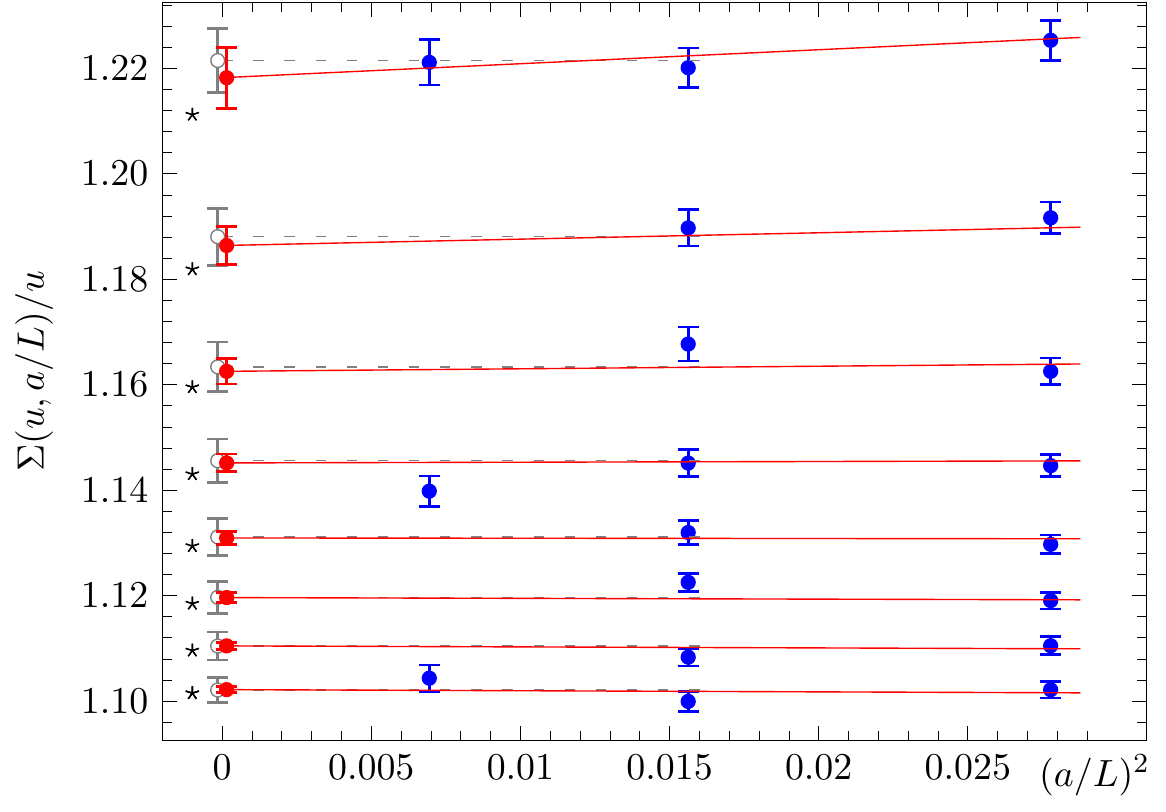}
  \caption{\label{f:cont}Continuum limit of the step scaling function
  $\Sigma^{(i)}(u,a/L)/u$ with $i=2$ loop improvement. As an 
  illustration a constant ($n_\rho =0$, dashed, fit G) and  a linear 
   ($n_\rho =2$, fit C) continuum extrapolation is shown. 
   Continuum extrapolated results include the errors due to $\ct$ and $\cttilde$ (cf.~text).
   The $\star$-symbols 
   show the perturbative $\sigma$ computed from the three-loop 
   $\betapert$.}
\end{figure}

As the residual linear $a$ effects are treated as an uncertainty, we can proceed
with continuum extrapolations linear in $a^2$. First we look at the data in \fig{f:cont}.
They are statistically compatible with having no 
$a$-effects for $L/a \geq 6$; for $\nf=0$ this was found
with similar precision for $L/a\geq 5$ (see Fig.~3 
of \cite{Bode:2001jv}).

Both the continuum limit of the \SSF\ 
and its cutoff effects are smooth functions of the coupling.
This motivates global fits of the form
\begin{eqnarray}
    \Sigma_\nu^{(i)}(u,a/L) &=& \sigma_\nu(u) + \rho_\nu^{(i)}(u) \, (a/L)^2 \, ,
\end{eqnarray}
where $i$ is the order of PT to which cutoff effects are 
removed in \eq{e:sigimpr}.
We performed various such fits in order to assess the systematic 
errors which result from the assumptions made in the fit
functions.
We parameterize the cutoff effects by 
a polynomial in $u$, with the correct asymptotics for small 
$u$, 
\begin{eqnarray}
    \rho_\nu^{(i)}(u) &=& \sum_{k=1}^{n_\rho^{(i)}} \rho^{(i)}_{\nu,k} u^{i+1+k}\, ,
    \label{e:rho}
\end{eqnarray}
where the case of neglecting cutoff effects is covered 
by ${n_\rho^{(i)}}=0$.
The continuum step scaling function is naturally 
parameterized by a polynomial in $u$,
\begin{eqnarray}
   \sigma_\nu(u)= u + u^2\sum_{k=0}^3 s_k u^{k}\,.
\end{eqnarray}
Lower order coefficients are fixed to their 
known perturbative values while $s_3$ (``$n_c=1$'') or 
$s_2,\,s_3$  (``$n_c=2$'') are
fit parameters.
A selection of such fits are illustrated in \tab{t:Sigfits}.
Instead of the parameters of the
continuum \SSF\ the table shows directly the extracted $\Lswi
\Lambda$, where $\Lswi$ is defined through \eq{eq:gL0} and the
value $u_0=2.012$. Recalling \eq{eq:gbarnu} and using $\bar v =
0.1199(10)$ (see next section) we have
\begin{eqnarray}
  \gbarnu^2(1/\Lswi) = 2.012\, (1 - 0.1199(10) \times 2.012\,\nu)^{-1}\,.
  \label{e:gnuL0}
\end{eqnarray} 
Apart from the form of the fit, $\Lswi \Lambda$ depends on 
the value of $n$ where
\eq{e:LLpert} with $\beta_\nu=\betapert_\nu$ is used. 
Since we insert $\betapert_\nu$ at three-loop, the residual dependence
on the coupling is $\rmO(\alpha^2(1/L_n))$. 

\begin{figure}[t]
   \includegraphics*[width=0.9\linewidth]{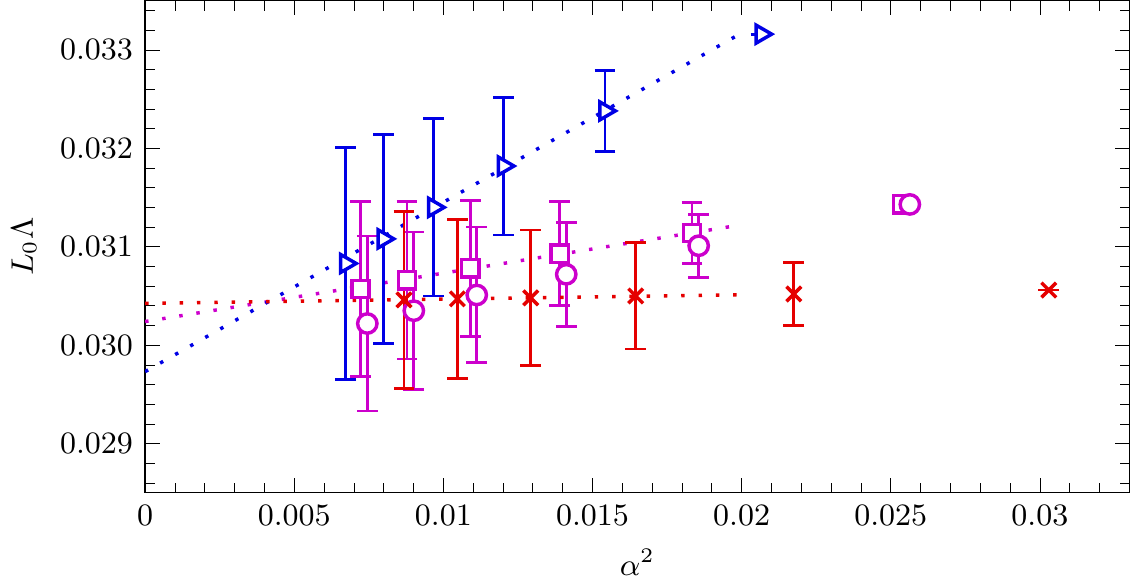}  
  \caption{\label{f:LLmax-extrap} 
   The dependence of the $\Lambda$-parameter on
   the coupling, $\alpha$.  
   From right to left, $n=0,1, \ldots,5$ steps of non-perturbative step-scaling are 
   performed to arrive at $\alpha(\mu)$ at $\mu=1/L_n$, before 
   using perturbative running. From top to bottom the different symbols 
  correspond to $\nu=-0.5,0,0.3$. We use 
  $i=1$ loop improved data and fit B, for $\nu=0$ we also show
  $i=2$, fit C. Dotted lines show linear dependence in $\alpha^2$ to
  guide the eye.
  \label{f:llplot}}
\end{figure}

The observed behavior, \fig{f:LLmax-extrap}, is consistent
with a dominatingly linear  dependence
of $\Lswi \Lambda$ on $\alpha^2(1/L_n)$. For $\nu=0$ the
slope is not very significant and for $\nu=0.3$ it about disappears, but for $\nu=-0.5$ it 
is quite large and outside errors. 

This suggests to perform alternative fits, where the continuum step scaling function is parameterized by an effective 
four-loop $\beta$-function, adding a term 
$b_3^\mathrm{eff}g^9$ to the known ones. 
The determined $\Lswi \Lambda$
are then automatically independent of $n$ and
 we include  $b_3^\mathrm{eff}$ instead
of $u_{n=4}$ in the table. For $\nu=-0.5$ the effective
fit value is  
larger than it should be in a well-behaved perturbative 
expansion. 

\begin{table}[h!]
\small\footnotesize
 \centering
\begin{ruledtabular}
\begin{tabular}{ccccccccccc}
 fit & $u_4$ & $i$ & $\left. \frac{L}{a}\right|_\mathrm{min}$ & $n_\rho^{(i)}$ & $n_c$ &
 $\Lswi \Lambda$ &  $b_3^\mathrm{eff}$ & $\chi^2$ & d.o.f. 
\\[-0.5ex]
 & & & &  &  & $\times 100$  &  $\times (4\pi)^4$ &  &  \\

\hline\\[-1ex]
   A &    1.193(4) &  0 &  6 &  2 &  1 &       3.04(\phantom{1}8) &  &  14.7 &  16  \\
   B &    1.194(4) &  1 &  6 &  2 &  1 &       3.07(\phantom{1}8) &  &  14.2 &  16  \\
   C &    1.193(5) &  2 &  6 &  2 &  1 &       3.03(\phantom{1}8) &  &  14.5 &  16  \\
   D &    1.192(7) &  2 &  6 &  2 &  2 &       3.03(13) &  &  14.5 &  15  \\
   E &  &  2 &  6 &  2 &  1 &      3.00(11) &         4(3) &  14.6 &  16  \\
  F &  &  2 &  8 &  1 &  1 &       3.01(11) &         4(3) &  12.7 &   \phantom{1}9  \\
   G &   1.191(11) &  2 &  8 &  0 &  2 &      3.02(20) &  &  13.0 &   \phantom{1}9  \\
   H &    &  1 &  6 &  2 &  1 &      3.04(10) &       3(3) &  14.1 & 16  \\
  \\[1ex]
fit &  $\nu$ & $i$ & $\left. \frac{L}{a}\right|_\mathrm{min}$ & $n_\rho^{(i)}$ & $n_c$ &
 $\Lswi \Lambda$ &  $b_{3,\nu}^\mathrm{eff}$ & $\chi^2$ &  d.o.f   
 \\[-0.5ex]
 & & & &  &  & $\times 100$  &  $\times (4\pi)^4$ &  &  \\
\hline\\[-1ex]
H  & $-$0.5 & 1 &  6 &  2 &  1 &       3.03(15) &  11(5) &  10.4 &  16  \\ 
H &   \phantom{$-$}0.3 & 1 &  6 &  2 &  1 &       3.04(10) &   \phantom{1}0(3) &  20.0 &  16  \\
\end{tabular}
\end{ruledtabular}
  \caption{Results for $\nu=0$ in the upper part. 
  }
\label{t:Sigfits}
\end{table}

We will come back to this issue shortly, but first we 
give our result for $\Lswi \Lambda$. We take the standard 
polynomial fit to $\sigma$ (for $\nu=0$) with
$\alpha_n \approx 0.1$ ($u_n\approx 1.2$). A typical perturbative 
error of size $\Delta(\Lambda L_n) = \alpha_n^2\,\Lambda L_n $
is then a factor 3 or 
more below our statistical errors. We quote (with  $\gbar^2(1/\Lswi) = 2.012$)
\begin{eqnarray} 
 \Lswi \Lambda =0.0303(8) \; \,   \to \;
 \Lswi \Lambda_\msbar^{(3)} = 0.0791(21)\,,
  \label{e:llresult2}
\end{eqnarray} 
with the known $\Lambda_\msbar/\Lambda$~\cite{Luscher:1993gh,Sint:1995ch}.
This is the result of fit~C. It is in perfect agreement with all variations of the 
global fit,  even with fit~G, which neglects all cutoff effects but 
uses only data with $L/a\geq 8$. It has a rather conservative error. If an even more conservative result is desired,
one may take the one of fit~D, $\Lswi \Lambda =0.0303(13)$.

\section{Accuracy of perturbation theory}
\begin{figure}[t]
   \includegraphics*[width=0.9\linewidth]{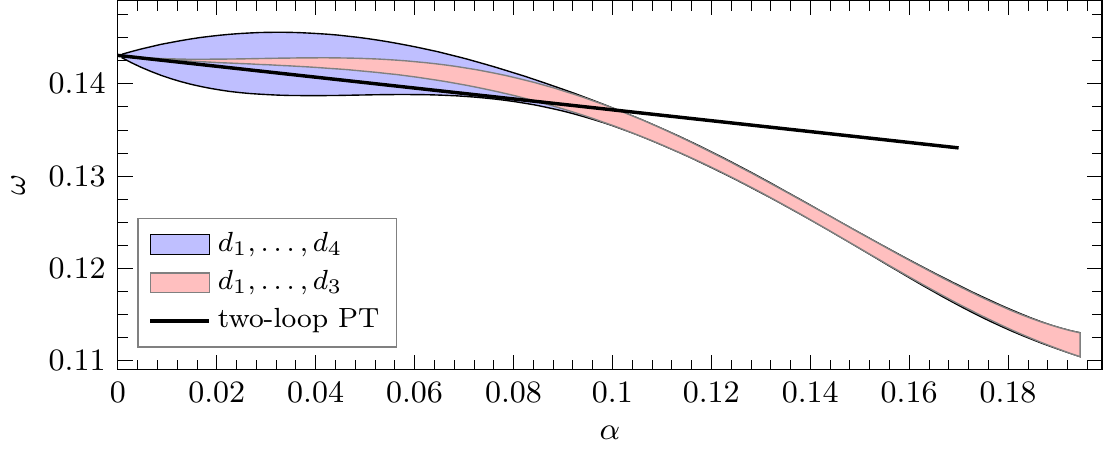}
  \caption{\label{f:omega} The function $\omega(\gbar^2)$ after continuum
  extrapolation, covering the $\pm1\sigma$ band of two fits described 
  in the text. 
  }
\end{figure}

While $b_{3,\nu}^\mathrm{eff}$ is large for $\nu=-0.5$, it does
have an error of around 50\%. A much better precision
can be achieved by considering directly the observable 
\begin{eqnarray}
  \omega(u) = \left.\vbar \right|_{\gbar^2(1/L)=u,m=0} 
  = v_1 + v_2 u + \rmO(u^2) \,,
  \label{e:omega}
\end{eqnarray}
with coefficients 
{  $v_1=0.14307,\,v_2=-0.004693$ \cite{DellaMorte:2004bc}}.
In contrast to the \SSF\, $\omega(u)$  does not require pairs of lattices,
so that the continuum extrapolation can be performed using
data for the entire range of lattice sizes $L/a=6,8,10,12,16,24$.
Improvement and fits for obtaining the continuum limit are carried out
in analogy to those of $\Sigma_\nu$. 
\Fig{f:omega} shows the result of two different fits
with 
fit parameters $d_k$ in $\omega(u)=v_1+v_2u+d_1u^2+d_2u^3+d_3u^4$ 
and in $\omega(u)=v_1+d_1u^1+d_2u^2+d_3u^3+d_4u^4$. 
The overall band of the two fits may be taken as 
a safe estimate of the continuum limit. 
As an example we find $\omega(2.012)=0.1199(10)$ for both 
fits, leading to \eq{e:gnuL0}. 
In the above analysis we did not use
data with $L/a=6$. Including them yields only tiny changes
and excellent $\chi^2$ values. 
 
A good measure of the deviation from two-loop perturbation 
theory is
\begin{eqnarray}
   (\omega(\gbar^2) - v_1 - v_2 \gbar^2 )/v_1 = -3.7(2) \,\alpha^2 
   \label{e:vbar3eff}
\end{eqnarray}
at $ \alpha=0.19$.
It is quite large and statistically significant
beyond any doubt. If one 
attempts to describe this by perturbation theory,
the  
three-loop coefficient $v_3$ has to be too
large for perturbation theory to be trustworthy 
at $\alpha=0.2$.
{  Again, we} come to the conclusion that 
$\alpha\approx 0.1$
needs to be reached non-perturbatively before perturbation theory becomes
accurate.


\section{Summary and Conclusions}

Our chosen definition of $\alpha_s(\mu)$ allows us to
compute it with very good precision
through lattice Monte Carlo simulations.
 In particular we have 
controlled the errors due to the discretisation of the theory
also at large $\mu$. Known non-perturbative corrections are parametrically 
very small: $\rmO(\rme^{-2.6/\alpha})$. In other words we have an excellent scheme to test the accuracy of PT in a given 
region of $\alpha$.

In fact, we have a family of schemes, depending on $\nu$. 
For small positive $\nu$, the couplings follow perturbation theory very closely in the full investigated range 
$0.1 \leq \alpha \leq 0.2$ as illustrated by the flatness 
of $\Lambda$ in \fig{f:llplot} extracted from \eq{e:LLpert}
with the three-loop $\beta$-function.

However, for negative $\nu$, e.g. $\nu=-0.5$, values of $\alpha$ 
just below 0.2 are not small enough to confirm 
perturbative behaviour. The observable $\vbar$,  \fig{f:omega}, shows
that the $\alpha$-dependence seen in \fig{f:llplot} is not just a statistical fluctuation. We could
take the continuum limit of $\vbar$ with very high precision and \eq{e:vbar3eff}
shows a clear deviation from the known perturbative 
terms, corresponding to $\lb=3$, at $\alpha=0.19$. 

We conclude that it is essential to  reach 
$\alpha=0.1$ in order to be able to achieve a precision
around $3\%$ for the $\Lambda$-parameter. Fortunately
we have access to that region and can quote such an accuracy in
\eq{e:llresult2}. 
While of course schemes exist where three-loop 
running holds accurately down to smaller energies -- for 
example $\nu=0.3$ 
produces flatness in \fig{f:llplot} 
as far as we can tell --
to know whether a chosen scheme possesses this property
is difficult unless one has 
control also over the $\alpha\approx 0.1$ region.
Once that is achieved larger $\alpha$ 
are not much needed any more. 

What we reported in this letter is part of our 
determination of a precise value for $\Lambda_\msbar$.
As our next step, we will soon
connect $\Lswi$ to the decay constants of pion and kaon,
as explained above and in \cite{Brida:2015gqj}.

\vspace{8mm}

\begin{acknowledgments}
We thank our colleagues in the ALPHA collaboration,
in particular M.~Bruno, C.~Pena, S.~Schaefer, H.~Simma and U.~Wolff for many useful discussions.
We thank U.~Wolff for a critical reading of the manuscript. 
We would also like to show our gratitude to 
S. Schaefer and H. Simma for their invaluable contribution in adapting the \texttt{openQCD} code.
 We thank the
computer centres at HLRN (bep00040) and NIC at DESY, Zeuthen for providing computing resources and support. S.S.
acknowledges support by SFI under grant 11/RFP/PHY3218.
 P.F. acknowledges financial support from the Spanish MINECO’s “Centro de 
Excelencia Severo Ochoa” Programme under grant SEV-2012-0249, as well as from 
the MCINN grant FPA2012-31686.
This work is based on previous work \cite{Sommer:2015kza} supported strongly by the Deutsche Forschungsgemeinschaft in the SFB/TR~09. 
\end{acknowledgments}
\vspace{15mm}

\section{Supplementary material}

Here we add some details on the Symanzik improvement of the action and
perturbative improvement of the \SSF~\cite{deDivitiis:1994yz}. In particular 
we discuss how the uncertainties in improvement coefficients are handled.

\subsubsubsection{Improvement of the action}

Apart from the bulk $\Oa$ improvement term, 
the complete removal of linear (in $a$) discretization
errors requires a boundary improvement coefficient $\ct$ 
in the gluon action \cite{Luscher:1992an} 
and a coefficient $\cttil$ in the fermion action \cite{Luscher:1996sc}. 
Regarding the former, the known two-loop accuracy
\cite{Bode:1999sm}, 
\begin{eqnarray}
 \ct &=& 1 + g_0^2 (-0.0890  + 0.019141 \nf) \nonumber \\ &&
         +\, g_0^4 (-0.0294 + 0.002 \nf + 0.000(1)\nf^2)\,,
\end{eqnarray}
is expected to be sufficient~\cite{Bode:2001jv} since we
are in the weak bare coupling region. 
Still, we propagate the small deficit of $\Oa$-improvement into our errors. As an 
uncertainty in $\ct$ we use the full two-loop term,
adding $\Delta^{\ct} \Sigma = 0.0234g_0^4|\partial_{\ct}\Sigma|$ in quadrature to the statistical errors.  
The derivative $\partial_{\ct}\Sigma$ was
obtained from a numerical estimate of
$\delta_b(\gbar^2) \equiv \frac{L}{2a} \frac{1}{\gbar^2}\partial_{\ct} \gbar^2$, namely
 $\delta_b(2.02)=-2.15(5) $
 with negligible dependence on $L/a$ and $\nu$.
 For this purpose
 we performed simulations on $L/a=6,8$ lattices
 with various values of $\ct$.   
Combined with $\delta_b(0)=-1$ we then use the interpolation 
$\delta_b(u) = - (1 + 0.57(3)\,u)$
and set $\partial_{\ct} \Sigma(u,a/L) = -(a/L) \,u\, \delta_b(u)$.

Similarly, for the coefficient 
$\cttil(g_0) = 1  -0.01795 g_0^2 + {\rm O}(g_0^4)$ \cite{Luscher:1996vw} we use
the full one-loop term as an error estimate.
Its effect is much smaller than the
one of the uncertainty in $\ct$, since it
contributes only through quark loops. 

These errors are responsible for around
30\% of the uncertainties of some of our central results in TABLE I.
Throughout error propagation is carried out
as detailed in \cite{Tekin:2010mm}.

\subsubsubsection{Perturbative improvement of the step scaling functions}

In addition, 
we can improve the 
observables to a given order $i$ in perturbation theory but to all orders
in $a$ via
\begin{equation}
  \Sigma^{(i)}(u, a/L) = \frac{\Sigma(u, a/L)}{1 + \sum_{k=1}^i\delta_k(a/L)\,u^k}\,,
  \label{e:sigimpr}
\end{equation}
with $\delta_1,\delta_2$ known~\cite{Bode:1999sm}.

\bibliographystyle{apsrev4-1} 
\bibliography{gbarlett}

\end{document}